\documentclass[conference]{IEEEtran}
\usepackage{cite}
\usepackage{amsmath,amssymb,amsfonts}
\usepackage{algorithmic}
\usepackage{graphicx}
\usepackage{textcomp}
\usepackage{xcolor}
\usepackage{comment}
\usepackage[nolist,nohyperlinks]{acronym}
\usepackage{multirow}
\usepackage[hidelinks]{hyperref}
\usepackage{url}

\usepackage[font=footnotesize]{subcaption}
\usepackage[font=footnotesize]{caption}

\newif\ifaccepted
\begin{document}
\newcommand\redbf[1]{\textcolor{red}{\textbf{#1}}}
\acrodef{DPD}{Digital Predistortion}
\acrodef{NN}{Neural Network}
\acrodef{flops}[FLOPs]{FLoating-point-OPerations}
\acrodef{LASSO}{Least Absolute Shrinkage and Selection Operator}
\acrodef{MRMR}{Minimum Redundancy Maximum Relevance}
\acrodef{DUT}{Device-Under-Test}
\acrodef{MP}{Memory Polynomial}
\acrodef{GMP}{Generalized Memory Polynomial}

\title{Low Complexity Neural Network Digital Predistortion of Wideband Power Amplifiers through Feature Selection}

\author{\IEEEauthorblockN{Cel Thys, Rodney Martinez Alonso, Ali H. Alsarraf, Dominique Schreurs, Sofie Pollin}
\IEEEauthorblockA{WaveCoRE, Department of Electrical Engineering, Katholieke Universiteit Leuven, Belgium}
}
\maketitle

\begin{abstract}
Due to the continuous increase in communication bandwidth and the use of highly efficient yet nonlinear power amplifiers, Digital Predistortion (DPD) algorithms are becoming increasingly complex. In particular, neural network (NN) based DPD approaches using Phase-Normalized NN architectures often incur substantially higher computational costs than widely deployed polynomial-based methods, such as the Memory Polynomial (MP) and Generalized Memory Polynomial (GMP) models. To bridge this gap between research performance and practical implementation, we propose a low-complexity Feature Selection NN DPD architecture. The proposed method employs an offline feature-engineering pipeline based on the Least Absolute Shrinkage and Selection Operator (LASSO) and the Minimum Redundancy Maximum Relevance (MRMR) algorithm to construct a compact and informative input representation. Using measured wideband FR3 power amplifier datasets that are publicly released with this work, we demonstrate up to 30\% reduction in computational complexity while maintaining comparable linearization performance.
\end{abstract}


\section{Introduction}\label{sec:introduction}
The evolution toward sixth-generation (6G) wireless systems is driving the exploration of new spectrum allocations beyond the congested sub-6\,GHz and millimeter-wave (mmWave) bands. The upper mid-band, also known as Frequency Range~3 (FR3) and spanning roughly 7--24\,GHz, has emerged as a promising candidate for 6G due to its favorable balance between coverage and available bandwidth~\cite{cui_6g_2025}. A key challenge in this band is power efficiency: to meet demanding link budgets at these frequencies, power amplifiers (PAs) must operate close to saturation, where nonlinear distortion is most pronounced. \ac{DPD} is the standard technique to linearize PAs in the digital domain, allowing efficient operation without sacrificing spectral purity~\cite{becerra_digital_2026}. As signal bandwidths continue to grow, the memory effects of wideband PAs become increasingly significant, requiring more complex DPD models.

Traditionally, DPD has been implemented using truncated Volterra series, of which the \ac{MP} and \ac{GMP}~\cite{morgan_generalized_2006} are the most widely deployed examples. These models are attractive in practice because they are linear in their parameters, admitting closed-form least-squares solutions with low complexity. However, they rely on basis functions that become highly correlated for high nonlinearity orders or high memory depths, limiting the scalability of Volterra models~\cite{liu_dynamic_2004,hongyo_2019_DNNDPD}. Indeed, for wideband signals with high peak-to-average power ratio (PAPR) and high modulation order, higher-order Volterra expansions are numerically ill-conditioned, capping the achievable linearization performance. \ac{NN} based DPD has been widely investigated as a means to overcome these limitations~\cite{khan_next-gen_2026}. By learning the predistortion from data rather than relying on a polynomial basis, NN models can linearize complex PA behavior not captured by Volterra representations~\cite{hongyo_2019_DNNDPD, wu_residual_2020, fischer-buhner_phase-normalized_2023, wu_opendpd_2024}. The real-valued time-delay \ac{NN}~\cite{liu_dynamic_2004} and its derivatives, such as residual architectures~\cite{wu_residual_2020} and the phase-normalized neural network (PNN)~\cite{fischer-buhner_phase-normalized_2023}, have demonstrated state-of-the-art linearization performance.

Despite this performance advantage, the adoption of NN-based DPD in practical implementation remains limited. The core obstacle is computational complexity~\cite{khan_next-gen_2026}: NN models capable of outperforming Volterra series typically require a significantly larger number of operations per sample, with implementations reported in the literature commonly requiring thousands of \ac{flops} per IQ sample~\cite{khan_next-gen_2026}. This stands in contrast to the few hundred \ac{flops} typical of a \ac{MP} or \ac{GMP} model~\cite{tehrani_2010_complexity}, and poses a serious challenge for real-time deployment in resource-constrained digital frontend hardware.

In this work, we improve the complexity-performance tradeoff of NN DPD models by targeting the input features. The main contribution is a feature engineering pipeline that combines the \ac{LASSO} regularization~\cite{wisell_behavioral_2008, barry_comparison_2021} and \ac{MRMR} ranking~\cite{peng_feature_2005} to construct a compact, maximally informative, and non-redundant input feature set from the same Volterra features used in \ac{GMP} models. The resulting Feature Selection NN-DPD achieves an improved complexity-performance tradeoff, matching the linearization accuracy of a PNN while reducing its computational cost by up to 30\%. We validate the approach using measurements on two 15\,GHz devices with 100\,MHz wideband signals. To support reproducible research and open  benchmarking~\cite{wu_opendpd_2024}, the measured FR3 PA datasets used in this work are publicly available~\cite{thys_fr3_2026}.

The outline of the work is as follows: in Section~\ref{sec:system-model}, we provide an overview of the considered DPD system model and the metrics used for DPD performance evaluation. Section~\ref{sec:feat-nn} describes the proposed feature engineering pipeline and the feature-based NN architecture. The analysis of the performance complexity tradeoff for DPD algorithms is presented in Section~\ref{sec:results}, including measured linearization results. We end our work with the main conclusion in Section~\ref{sec:conclusion}.

\section{System model}\label{sec:system-model}
In this section, we introduce the \ac{DPD} problem formulation and the key metrics used to compare the performance and complexity of \ac{DPD} algorithms.

The operation of the \ac{DPD} system is as follows: an oversampled communication waveform $x(n)$ is predistorted by a nonlinear \ac{DPD} model, denoted by $f_{PD}$, to counteract the power amplifier (PA) nonlinearity. After predistortion, the waveform $z(n)$ is sent to the RF frontend, including the PA, after which an RF coupler provides a feedback signal $y(n)$. 
\begin{IEEEeqnarray}{rCl}
    z(n) & = & f_{PD}\{x(n)\},\\
    y(n) & = & f_{PA}\{z(n)\}.
\end{IEEEeqnarray}

We implement an indirect learning approach to DPD training, where measured PA response data are used to identify an inverse model that maps the PA output back to its input~\cite{eun_new_1997}, essentially performing post-distortion. Assuming that the PA behavior is sufficiently invertible within the operating region of interest, the identified post-distorter can be used as a predistorter by placing it before the PA. The predistortion function is implemented using truncated Volterra series or neural network models. Each model has coefficients $\theta$, which are optimized during the identification (or training) stage to minimize the mean square error (MSE):
\begin{equation}
    \min_{\theta} \frac{1}{N} \sum_{n=1}^{N} \left|x(n) - f_{PD,\theta}\{y(n)\} \right|^2 \label{eq:optimisation}
\end{equation}
This minimization problem is solved in closed form using least squares for Volterra-series models, whereas for NN-based models it is solved iteratively using stochastic gradient descent.

\subsection{Measurement system}
\begin{figure}[htbp]
\includegraphics[width=\linewidth, height=0.5\linewidth, keepaspectratio]{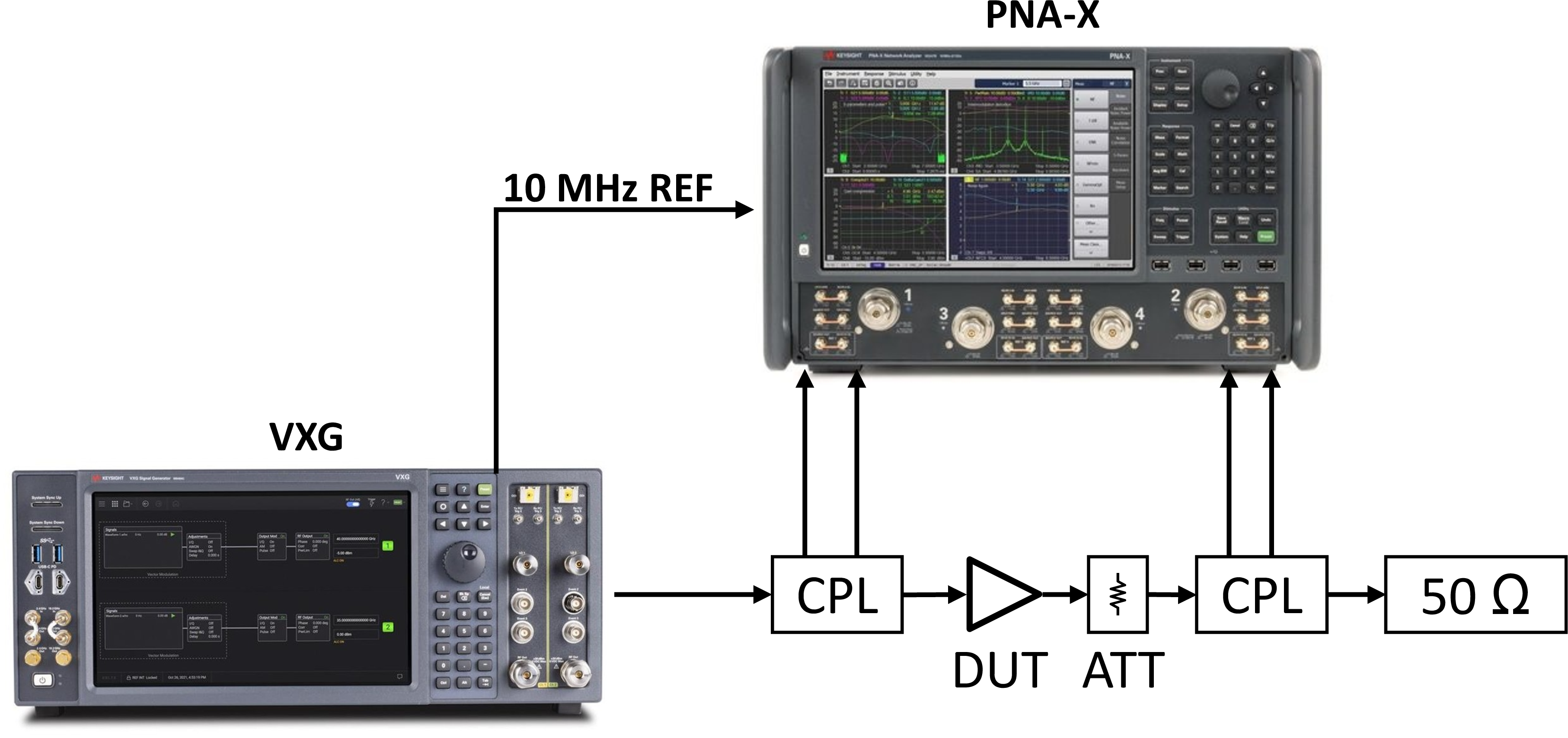}
\caption{Block diagram of the VNA-based measurement setup for DPD.}
\label{fig:meas_setup}
\vspace{-3mm}
\end{figure}

The setup for wideband modulated measurements~\cite{dunsmore_evolution_2024}, pictured in Fig.~\ref{fig:meas_setup}, consists of: a microwave vector signal generator (VXG) and a vector network analyzer (PNA-X) for signal generation and capture, bidirectional couplers, the \ac{DUT}, an attenuator (ATT) after the DUT to limit output power, and a 50-Ohm load.

Each experiment employs a single-carrier input waveform with 100 MHz 64-QAM modulated data, root raised cosine pulse shaping with factor $\alpha=0.35$ and sampling at 1.28\,GHz. Each dataset is split into training, validation, and test captures with proportions of 33-33-33 \%, respectively. DPD algorithm training and validation is performed in MATLAB, while the best-performing models are validated on the actual measurement setup. The measured FR3 PA response datasets are available at~\cite{thys_fr3_2026}.

\subsection{Performance metrics}\label{sec:performance-metrics}
After model identification, we use the normalized mean square error (NMSE) \eqref{eq:nmse} on a validation dataset that was not used for model identification as the performance metric to compare predistorters. Here $x(n)$ represents the true PA input and $z(n)$ represents the NN model prediction:
\begin{equation}
   NMSE = 10 \cdot \log_{10} \left\{ \frac{\frac{1}{N}\sum\limits_{n=1}^{N}|z(n)-x(n)|^2}{\frac{1}{N}\sum\limits_{n=1}^{N}|x(n)|^2} \right\} . \label{eq:nmse}
\end{equation}

For the best-performing models, we measure the PA response to the DPD waveform and report the error vector magnitude (EVM) and the adjacent channel leakage ratio (ACLR). Note that the validation NMSE can be calculated for each trained model in simulation, while the EVM and ACLR are measurement metrics reported by the testbench equipment:
\begin{equation}
{EVM} = 100 \cdot \sqrt{\frac{\frac{1}{N}\sum\limits_{n=1}^{N}|y(n)-x(n)|^2}{\frac{1}{N}\sum\limits_{n=1}^{N}|x(n)|^2}} , \label{eq:evm}
\end{equation}

\begin{equation}
ACLR = 10 \cdot \log_{10} \left( \frac{P_{adj}}{P_{main}} \right) . \label{eq:aclr}
\end{equation}

In the equations above, $y(n)$, the output of the combined DPD-PA system, is normalized by the mean gain of the PA, while $P_{adj}$, $P_{main}$ represent adjacent channel and main channel power, respectively. The measured EVM includes symbol recovery and equalization filtering as implemented in the VNA instrument.

\subsection{DPD algorithmic complexity} \label{sec:complexity}
In addition to the performance metrics mentioned above, we evaluate the DPD algorithmic complexity. Similar to \cite{tehrani_2010_complexity}, we focus on comparing the runtime complexity of DPD algorithms, as this is the most critical for real-time implementation. The complexity of offline model training or online adaptation is not considered, as this functionality does not need to run in real time.

Analysis of DPD complexity is typically limited to considering the number of real-valued parameters in the DPD model \cite{morgan_generalized_2006} as an indicator of its memory cost. We use \ac{flops} as a more accurate complexity metric and compare different DPD algorithms in an implementation-agnostic way~\cite{tehrani_2010_complexity}. In this way, we compare the number of computations to be performed by a hardware platform to execute the DPD algorithm, without considering deeper implementation details such as parallelism or pipelining.

We introduce the symbol $C$ for complexity in \ac{flops}~\cite{tehrani_2010_complexity}, with a subscript indicating a specific algorithm; e.g., $C_{MP}$ denotes the complexity of a Memory Polynomial DPD model. To calculate $C$ for specific DPD models, we must identify all operations performed in the model and their associated \ac{flops} cost. For Volterra models, we follow the approach in \cite{tehrani_2010_complexity}, splitting the complexity cost into basis construction and filtering: $ C = C_B + C_F$. For NN models, we develop our own approach by summing the costs of input feature generation, phase normalization preprocessing, the hyperbolic tangent nonlinear activation, and fully connected layer matrix multiplication and bias addition.

\begin{table}[htbp]
\caption{Computational complexity expressions in FLOPs/sample.}
\label{table:complexity}
\centering
\setlength{\tabcolsep}{2pt}
\begin{tabular}{l l l}
\hline
Model & Component & Complexity \\
\hline

\multirow{3}{*}{MP}
& Basis generation
& $C_{MP,B}=3+7+2(P-1)$ \\
& Filtering
& $C_{MP,F}=8MP-2$ \\
& Total
& $C_{MP}=C_{MP,B}+C_{MP,F}$ \\
\hline

\multirow{3}{*}{GMP}
& Basis generation
& $C_{GMP,B}=3+7+2(K_a-1)+4(K_b-1)M_b$ \\
& Filtering
& $C_{GMP,F}=8L_a(K_a+2K_bM_b)-2$ \\
& Total
& $C_{GMP}=C_{GMP,B}+C_{GMP,F}$ \\
\hline

\multirow{3}{*}{NN}
& Phase norm.
& $C_{PN}(I)=6(I+1)$ \\
& Fully connected
& $C_{FC}(I,O)=2IO$ \\
& Tanh activation
& $C_{TANH}(I)=10I$ \\
\hline
\end{tabular}
\end{table}

This work considers two Volterra series approaches: the \ac{MP} model with memory depth $M$ and polynomial order $P$; and the \ac{GMP} model with memory depth $L_a=L_b=L_c$, polynomial order $K_a$ and $K_b=K_c$ and delay/advance memory $M_b=M_c$ (notation from \cite{morgan_generalized_2006}).  For both models, we use even and odd-order envelope terms; although using only odd orders would achieve lower complexity, the DPD performance was measured to be similar. The cost of basis construction $C_B$ accounts for the generation of envelope terms of the form $|x(n)|^p$, while discounting any terms that can be taken from a delay line. The cost of Volterra filtering $C_F$ is derived from the number of model parameters, which is $M \cdot P$ for the \ac{MP} model and $K_aL_a+2K_bM_bL_a$ for the \ac{GMP} model.

We consider two neural network approaches to DPD: the baseline PNN~\cite{fischer-buhner_phase-normalized_2023} and the proposed Feature Selection \ac{NN}. Complexity formulations for all three are listed in Table~\ref{table:complexity} above, where $I$ and $O$ represent the size of that layer's input and output. We count 6 operations for each complex multiplication in the phase-normalization operation, which is applied to all input features and also to the output of the neural network~\cite{fischer-buhner_phase-normalized_2023}. In accordance with low-complexity implementations of the Tanh activation, we set the number of operations for this layer to 10~\cite{beebe_tanh_1991}.

The tunable hyperparameters, including the number of neurons per layer and the number of hidden layers determine the total complexity of NN model inference. Another hyperparameter is $k$, the number of input features, as explained in Section~\ref{sec:feat-nn}. This is the input to the first hidden layer of the NN model. To enable comparison across different complexities, these model hyperparameters are tuned using Bayesian optimisation~\cite{snoek_practical_2012}, and in Section~\ref{sec:results} we report the Pareto front of the best-performing models.

% this 2 column figure is going on the next page
\begin{figure*}[tb]
\includegraphics[width=\linewidth, height=0.5\linewidth, keepaspectratio]{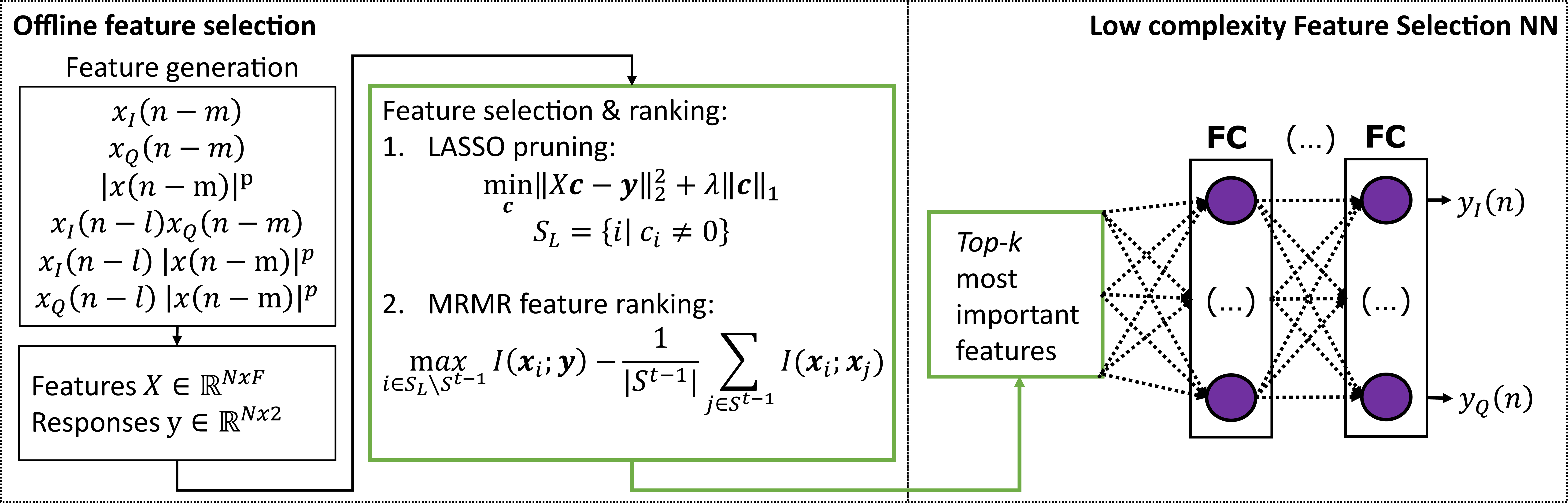}
\caption{Proposed feature selection approach to low complexity NN DPD.}
\vspace{-5mm}
\label{fig:feature-selection}
\end{figure*}

\section{Proposed low complexity NN-DPD}\label{sec:feat-nn}
The proposed approach, summarized in Fig.~\ref{fig:feature-selection}, decouples the DPD problem into two stages: an offline feature selection pipeline and a low-complexity neural network suitable for real-time inference. We hypothesize that state-of-the-art NN DPD algorithms are generally quite complex because they perform data-driven feature extraction during each inference. Our proposed method decouples feature extraction from inference, thereby reducing computational load.

The offline feature selection stage constructs a rich Volterra-inspired feature basis space and then applies a feature engineering process to identify the most informative and least redundant subset of features. The online stage consists of a compact real-valued neural network that maps the selected features to the predistorted output. Since the feature selection is performed once during model design and is not subject to real-time deadlines, its computational cost is not included in the DPD runtime complexity model introduced in Section~\ref{sec:complexity}, where only the top-\textit{k} feature generation and NN model inference are counted.

In this section, we adopt the conventional machine learning notation, where $x(n)$ is the NN model input and $y(n)$ denotes the NN target output. This corresponds to the notation for a PA behavioral modeling task, while under the indirect learning framework~\cite{eun_new_1997} introduced in Section~\ref{sec:system-model}, these signal roles are to be swapped such that the learned DPD model approximates the inverse PA characteristic.

\subsection{Feature Generation}
The starting point for the feature basis space is a truncated Volterra-like decomposition of the input signal $x(n)$. Because the neural network operates on real-valued inputs, the complex baseband signal is projected into its real and imaginary parts. For memory depth $M$ and maximum nonlinearity order $P$, each input sample contributes the following feature types:
\begin{align}
    &x_I(n-m), \quad x_Q(n-m) \label{eq:feat_linear} \\
    &|x(n-m)|^p \label{eq:feat_amp} \\
    &x_I(n-l)\,x_Q(n-m) \label{eq:feat_crossIQ}\\
    &x_I(n-l)\,|x(n-m)|^p, \quad
     x_Q(n-l)\,|x(n-m)|^p \label{eq:feat_cross}
\end{align}
where $m, l \in \{0,\ldots,M-1\}$ and $p \in \{1,3,\ldots,P\}$. The delay line
\eqref{eq:feat_linear} captures the linear memory of the \ac{DUT}, while the envelope
terms \eqref{eq:feat_amp} account for memoryless nonlinearity with odd order $p$. The cross terms \eqref{eq:feat_crossIQ} nonlinearly combine I/Q terms at different memory depths, while the terms \eqref{eq:feat_cross} combine instantaneous samples with envelope terms, similar to the \ac{GMP} structure~\cite{morgan_generalized_2006}. Together, these span a feature space with expressive power similar to that of a large \ac{GMP} model, with the important distinction that all features are real-valued scalars. Note that in a typical \ac{NN} DPD model, terms such as \eqref{eq:feat_crossIQ}, \eqref{eq:feat_cross} are not used as an input, as their construction would consume too much computational resources. The feature generation procedure yields a large input feature matrix $X \in \mathbb{R}^{N \times F}$, where the total number of candidate features $F$ scales quadratically in $M,P$. The corresponding response matrix $y \in \mathbb{R}^{N \times 2}$ holds the real and imaginary parts of the target signal. In the proposed feature engineering pipeline, the number of training samples $N$ is set to $15000$, while $M=200$ and $P=7$, leading to $F=321200$ features.

\subsection{LASSO Feature Selection}
A known limitation of high-order Volterra basis functions is their strong mutual correlation, which makes it difficult to assess individual feature relevance~\cite{khan_next-gen_2026, becerra_digital_2026}. While linear least-squares fitting in this setting is ill-conditioned, the \ac{LASSO} (or $\ell_1$ regularization) promotes sparsity and has relatively low complexity~\cite{barry_comparison_2021}. The \ac{LASSO} objective combines the standard least squares formulation with an $\ell_1$ sparsity penalty:
\begin{IEEEeqnarray}{rCl}
\hat{\mathbf{c}}
&=&
\arg\min_{\mathbf{c}\in\mathbb{R}^{F}}
\;
\|X\mathbf{c}-\mathbf{y}\|_2^2
+
\lambda\|\mathbf{c}\|_1
\label{eq:lasso}
\\
\mathcal{S}_{\mathrm L}
&=&
\{i : \hat c_i \neq 0\}.
\end{IEEEeqnarray}

Here $\hat{\mathbf{c}}$ is the sparse coefficient vector of the \ac{LASSO} model fit on the data and $\mathcal{S}_{\mathrm L}$ is the subset of selected features. The regularization parameter $\lambda$ is chosen via exhaustive search, aiming for approximately $2000$ surviving features after the LASSO stage. This stage significantly reduces the candidate feature space before feature ranking.

\subsection{MRMR Feature Ranking}
The goal of this stage is to rank the selected features in $\mathcal{S}_{\mathrm L}$ by order of importance, so that selecting the top $k$ features as inputs for the \ac{NN} will result in the most informative and least redundant features. We apply the \ac{MRMR}~\cite{peng_feature_2005} algorithm, which applies a greedy selection strategy, selecting in each iteration $t$ one optimal feature $i_t$ to add to the set $S^t$. Starting from $\mathcal{S}^0=\emptyset$, \ac{MRMR} iterates as follows:

\begin{IEEEeqnarray}{rCl}
i_t
&=&
\arg\max_{i\in\mathcal{S}_{\mathrm L}\setminus\mathcal{S}^{t-1}}
I(\textbf{x}_i;\textbf{y})
-
\frac{1}{|\mathcal{S}^{t-1}|}
\sum_{j\in\mathcal{S}^{t-1}}
I(\textbf{x}_i;\textbf{x}_j)
\label{eq:mrmr}
\\
\mathcal{S}^{t}
&=&
\mathcal{S}^{t-1}\cup\{i_t\}.
\end{IEEEeqnarray}
where $I(\textbf{x}_i;\textbf{y})$ denotes the mutual information between the feature vector $\textbf{x}_i$ and the output and $I(\textbf{x}_i;\textbf{x}_j)$ denotes the mutual information between features $i$ and $j$. The greedy solution produces a ranked list $i_t$, and the top $k$ features from this ranking are retained as the final input to the neural network. The hyperparameter $k$ directly controls the complexity-performance tradeoff and is treated as a model hyperparameter subject to the Bayesian optimization described in Section~\ref{sec:complexity}. The optimal range for $k$ is 100-200 features.

\subsection{Low-Complexity Feature Selection NN}
We adopt the residual PNN architecture~\cite{wu_residual_2020, fischer-buhner_phase-normalized_2023} as the network architecture, replacing its input with the feature vector of size $k$, while keeping the phase-normalization operation for I and Q inputs. The network produces outputs $y_I(n)$ and $y_Q(n)$, from which the complex postdistorter signal is constructed. By using a small $k$ relative to the full feature set $F$, the proposed pipeline can match or exceed the performance of an equivalently sized PNN while operating at a fraction of the computational cost.

\section{Results}  \label{sec:results}
We analyze the complexity-performance tradeoff for the baseline DPD algorithms and the proposed Feature Selection \ac{NN}, using measured data from two \ac{DUT}s that we make available in~\cite{thys_fr3_2026}. To approximate the true performance-complexity tradeoff for each of the considered DPD models, we run a hyperparameter search using Bayesian optimisation~\cite{snoek_practical_2012} and plot the convex hull of hyperparameter combinations that result in optimal performance at minimal complexity. Finally, we validate the proposed feature selection approach on the measurement system and report the EVM and ACLR performance.

\subsection{DPD modelling results}
\begin{figure}[htbp]
\includegraphics[width=\linewidth, keepaspectratio]{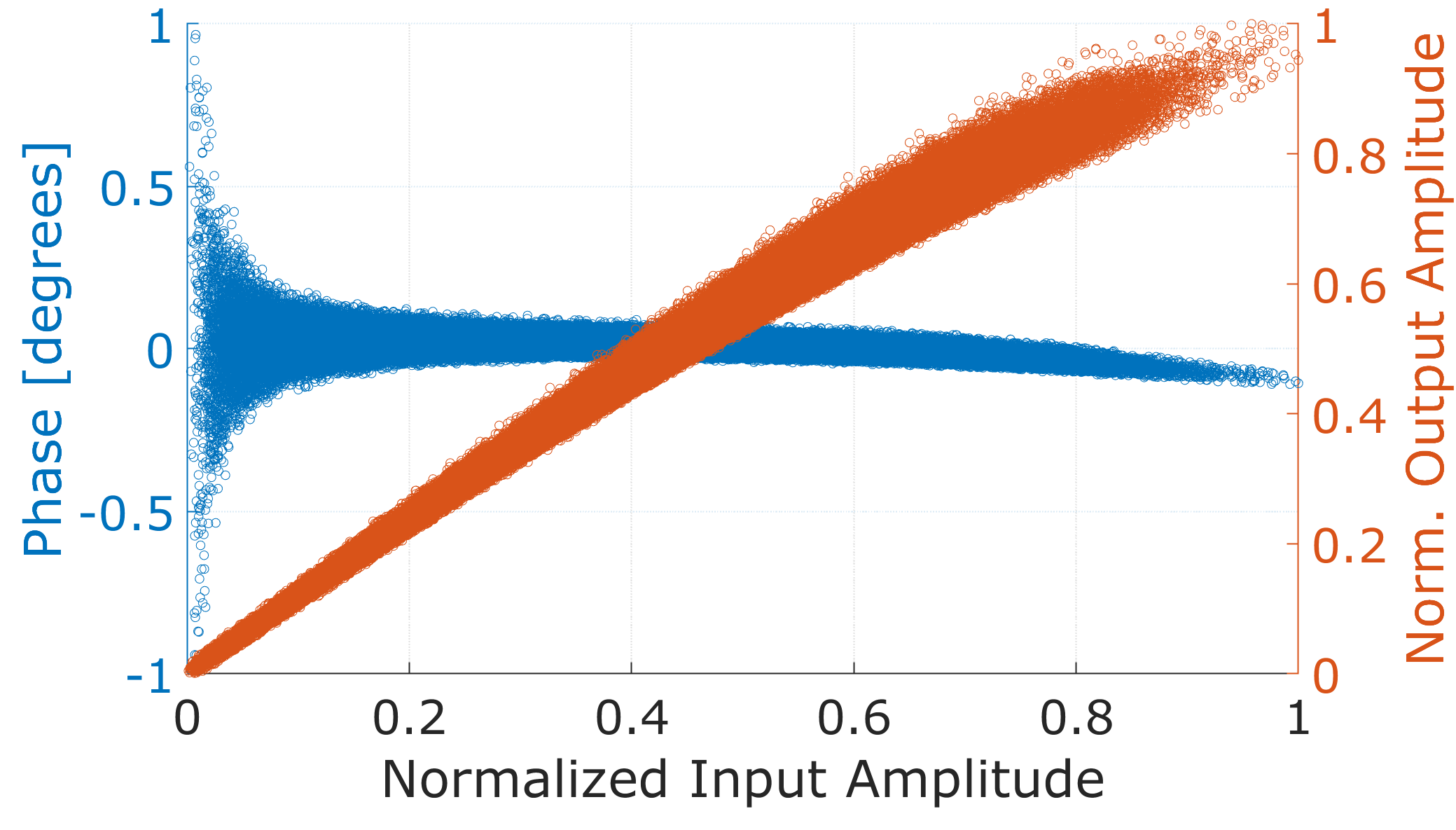} %height=0.5\linewidth, 
\caption{Measured AM/AM and AM/PM response of the 1st DUT.}
\vspace{-3mm}
\label{fig:lna_amam}
\end{figure}

Our first analysis focuses on DPD performance on a 15GHz GaAs amplifier (\textit{ADL9006}), which offers wideband amplification with a low gain of 15\,dB. Fig.~\ref{fig:lna_amam} shows the amplitude-amplitude (AM/AM) and amplitude-phase (AM/PM) response for this device at our chosen operation point of -2\,dBm input power and 10\,dB attenuation.

\begin{figure}[htbp] 
\includegraphics[width=\linewidth, keepaspectratio]{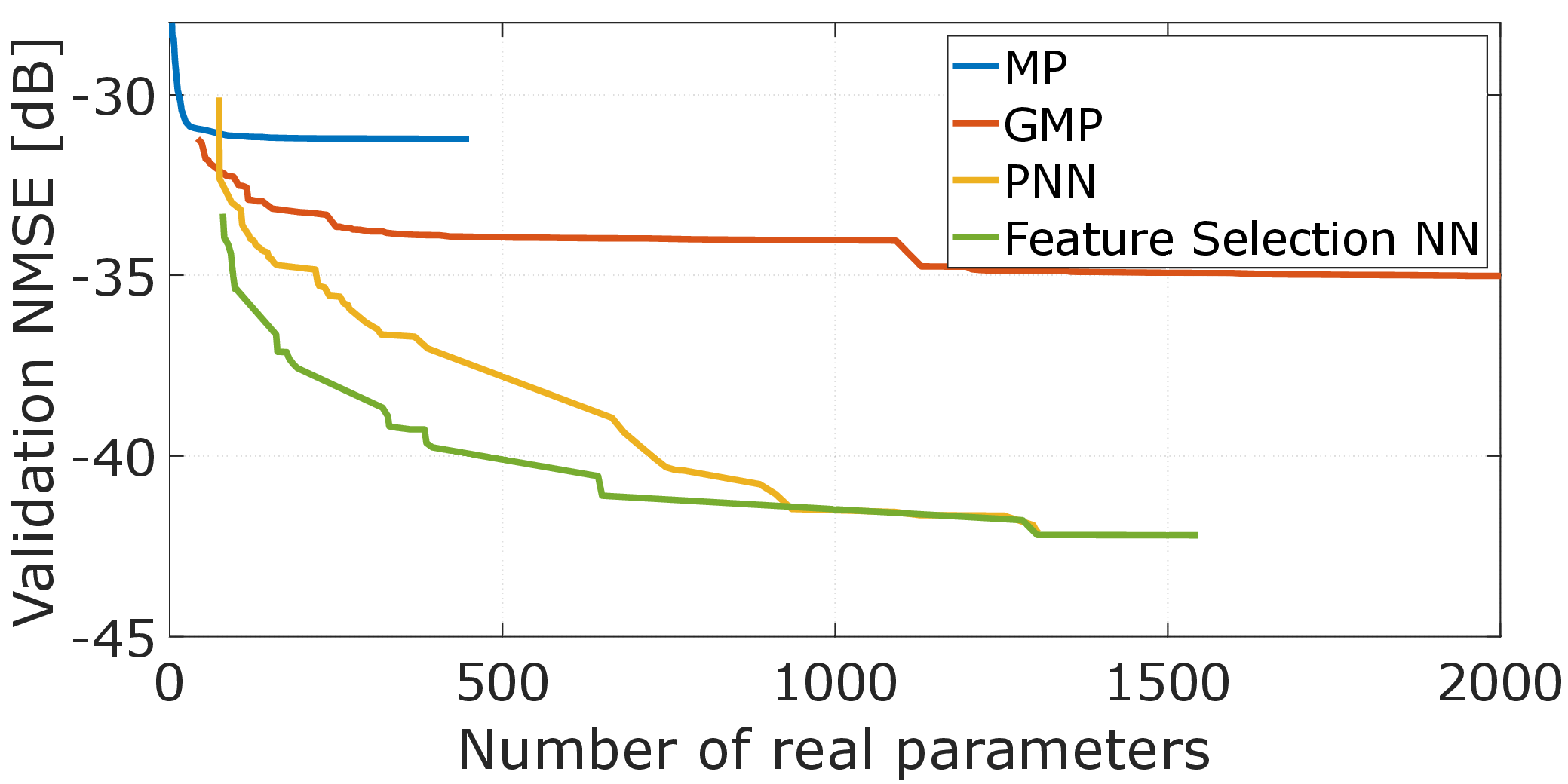} 
\caption{Validation NMSE as a function of DPD model size.} 
\vspace{-3mm}
\label{fig:lna_nmse_params} 
\end{figure}

Fig.~\ref{fig:lna_nmse_params} shows the DPD validation NMSE of different algorithms, as a function of the memory footprint in terms of the number of real-valued parameters. It highlights the remarkable performance of \ac{NN}-based \ac{DPD} algorithms, with both the baseline PNN model and the proposed approach achieving an NMSE of $-42$\,dB with 1600 parameters. The performance of the MP and GMP baselines stagnates quickly for relatively low number of coefficients.

\begin{figure}[htbp] 
\includegraphics[width=\linewidth, keepaspectratio]{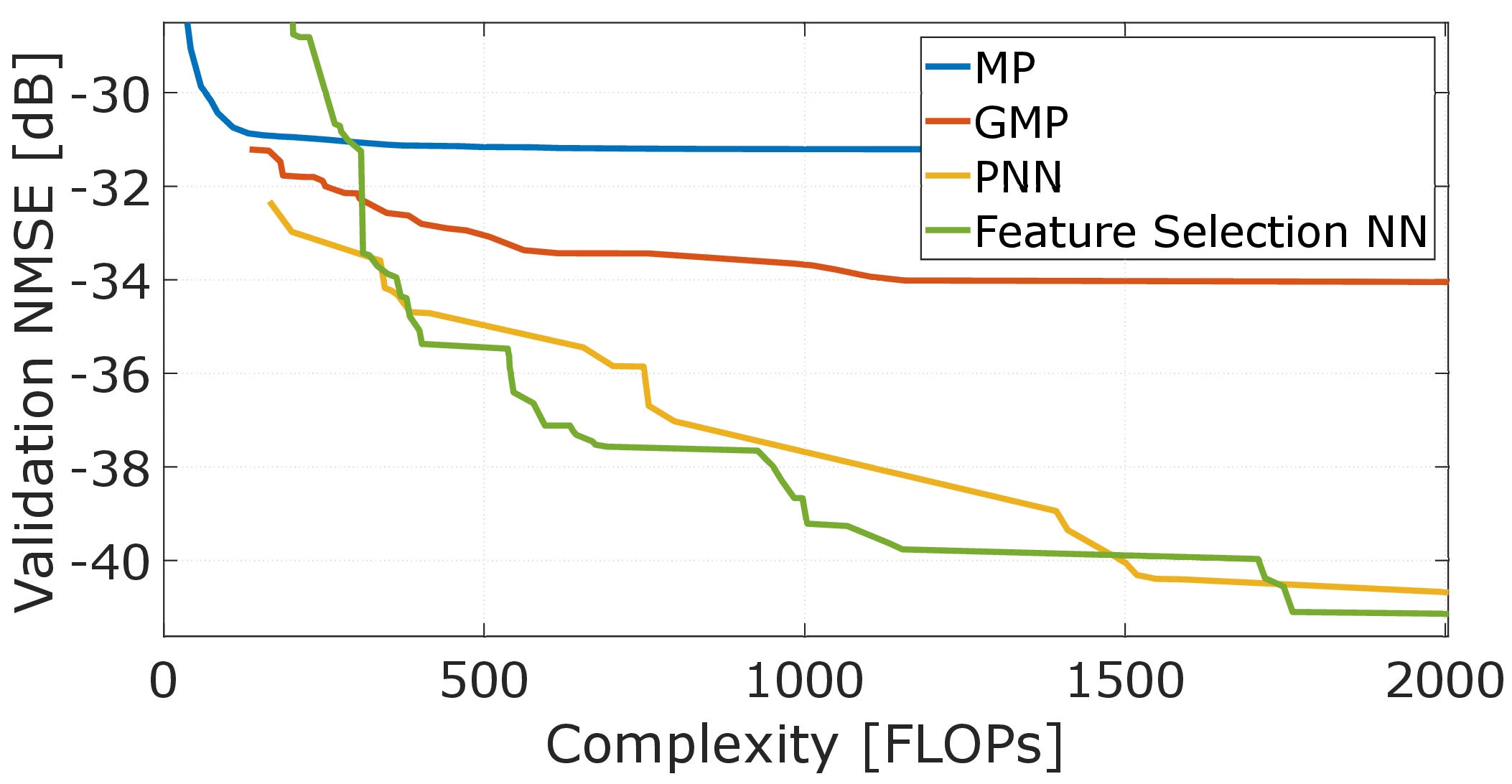} 
\caption{Validation NMSE as a function of DPD computational complexity.} 
\vspace{-2mm}
\label{fig:lna_nmse_flops} \end{figure}

Fig.~\ref{fig:lna_nmse_flops} shows the validation NMSE compared to DPD model computational complexity. It shows the complexity gap between the PNN baseline and the proposed Feature Selection NN. In fact, the proposed feature-based NN DPD model can significantly improve the tradeoff between NMSE and complexity, achieving $-37$\,dB NMSE with only 595 \ac{flops}. For similar NMSE performance, the PNN baseline requires 797 \ac{flops}, indicating the proposed approach offers an approximate 25\% improvement in \ac{flops}.

\begin{figure}[htbp]
\includegraphics[width=\linewidth, height=0.5\linewidth, keepaspectratio]{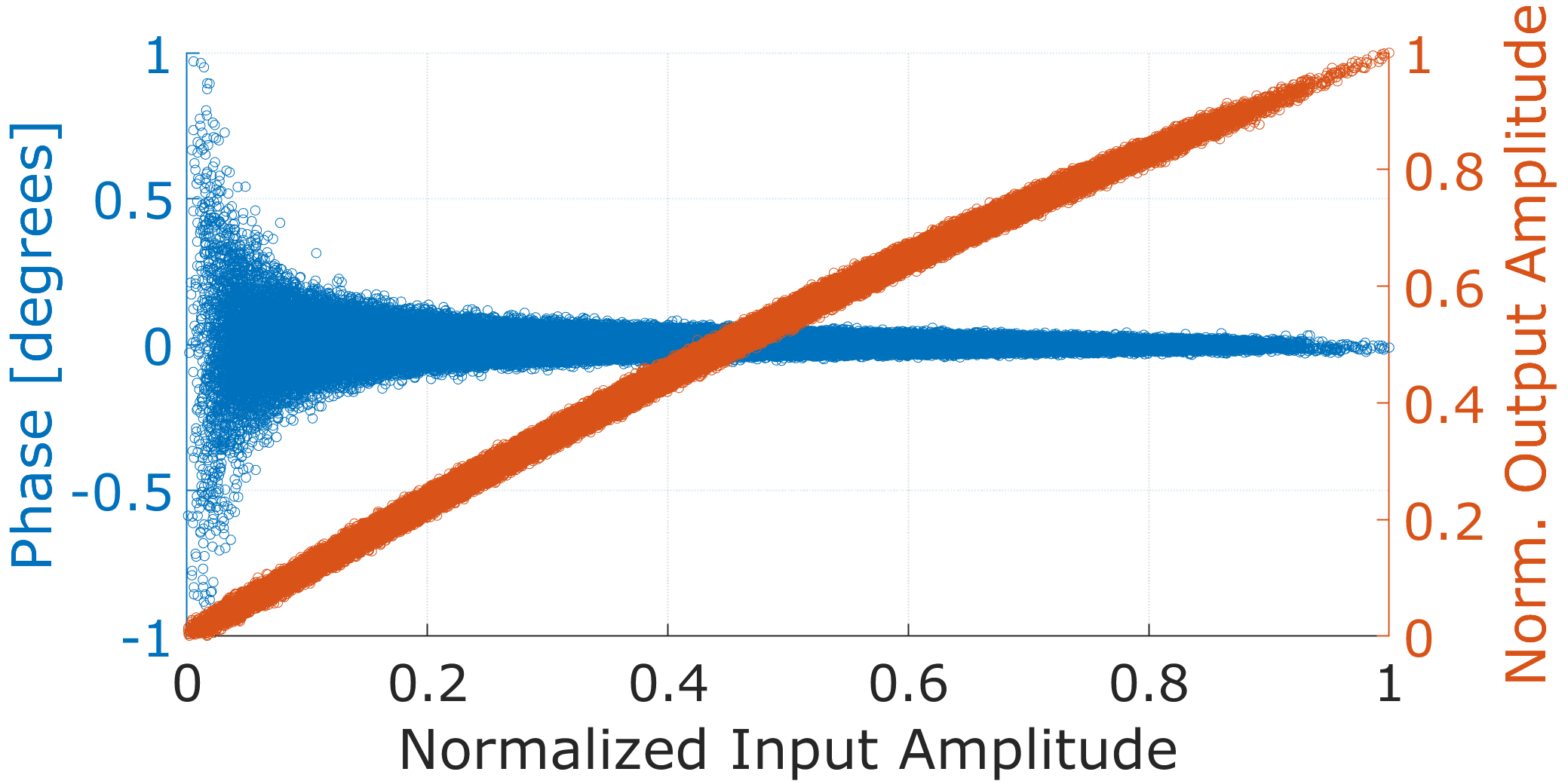}
\caption{Measured AM/AM and AM/PM response of the 2nd DUT.}
\vspace{-3mm}
\label{fig:tlpa_amam}
\end{figure}

Our next analysis concerns a high-gain PA (\textit{TLPA2G22G-43-43-HS}). This PA promises gain up to $53$\,dB over the frequency range of $2-22$\,GHz, sacrificing linearity in the process. Fig.~\ref{fig:tlpa_amam} shows the AM/AM and AM/PM responses for this device at $-20$\,dBm input power and 30\,dB output attenuation.

\begin{figure}[htbp]
\includegraphics[width=\linewidth, height=0.5\linewidth, keepaspectratio]{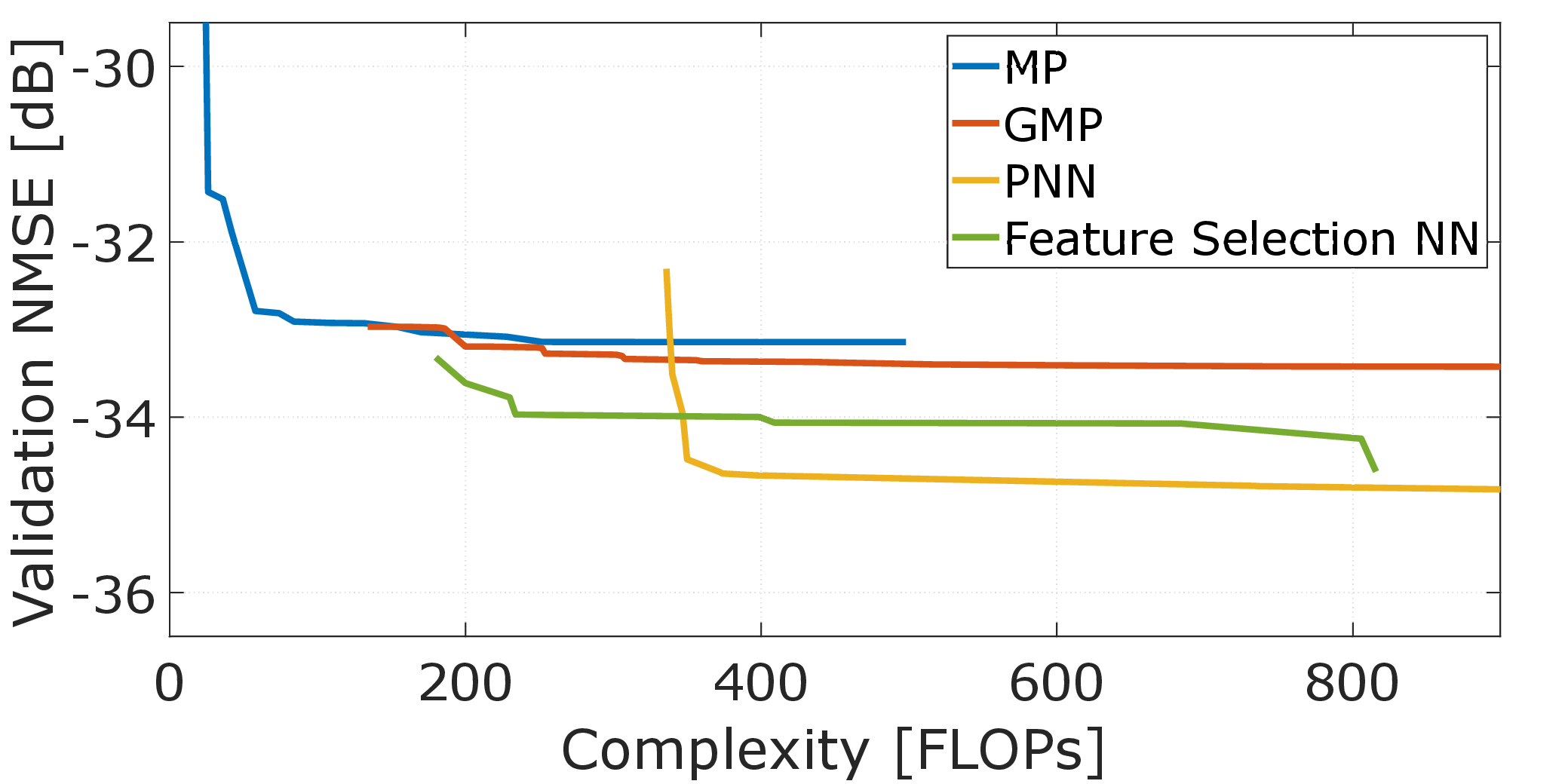}
\caption{Validation NMSE as a function of DPD computational complexity for the 2nd DUT.}
\vspace{-2mm}
\label{fig:tlpa_nmse_flops}
\end{figure}

Fig.~\ref{fig:tlpa_nmse_flops} shows the validation NMSE as a function of complexity for this challenging linearization task. The proposed feature selection NN improves the complexity-NMSE tradeoff in the low complexity region, achieving -34\,dB NMSE using only 234 \ac{flops}, which is a 32\% reduction compared to the PNN baseline.

\subsection{Measured DPD performance}
Several candidate models covering the NMSE/FLOP tradeoff were evaluated experimentally by generating predistorted waveforms and measuring the DPD performance on the first DUT (\textit{ADL9006}). The measured EVM and ACLR results are summarized in Table~\ref{table:measured_performance}, showing the \ac{flops}, the number of coefficients, and the performance metrics for 7 measured DPD models.

\begin{table}[!htbp]
\caption{Measured DPD performance on the 1st DUT. ACLR is reported as the worst case (maximum of upper and lower adjacent channels).}
\label{table:measured_performance}
\centering
\footnotesize
\setlength{\tabcolsep}{4pt}
\begin{tabular}{lcccccc}
\hline
\textbf{Model} & \textbf{FLOPs} & \textbf{Coef.} & \textbf{NMSE [dB]} & \textbf{EVM [\%]} & \textbf{ACLR [dBc]}\\
\hline
%\multicolumn{2}{l}{No DPD}
No DPD & --  & -- & -- & 2.5 & $-33.0857$ \\
\hline
\multirow{4}{*}{PNN}
&\textbf{360}&128&-32.7998&1.88&$-39.9457$\\
&787&287&-35.1755&1.54&$-41.0505$ \\
&1839&725&-37.4431&1.39&$-39.0286$ \\
&2491&911&-41.1065&1.39&$-41.4168$ \\
\hline
\multirow{3}{*}{Proposed}
&\textbf{407}&98&-35.3823&1.38&$\mathbf{-43.8113}$ \\
&939&344&-37.6726&1.24&$-41.2444$ \\
&1768&650&\textbf{-41.1615}&\textbf{1.08}&$-41.876$ \\
\hline
\end{tabular}
\end{table}

Table~\ref{table:measured_performance} shows that the proposed Feature Selection NN consistently achieves lower NMSE and EVM compared to a PNN model with the same complexity; for example it achieves 1.38\% EVM with only 407 \ac{flops}, with the baseline PNN achieving 1.88\%. The proposed Feature Selection NN achieves the lowest worst-case ACLR, up to 10dB lower than the amplifier without DPD and 3.8\,dB lower than a comparable PNN model.

\begin{figure}[!htbp]
\includegraphics[width=\linewidth, height=0.5\linewidth, keepaspectratio]{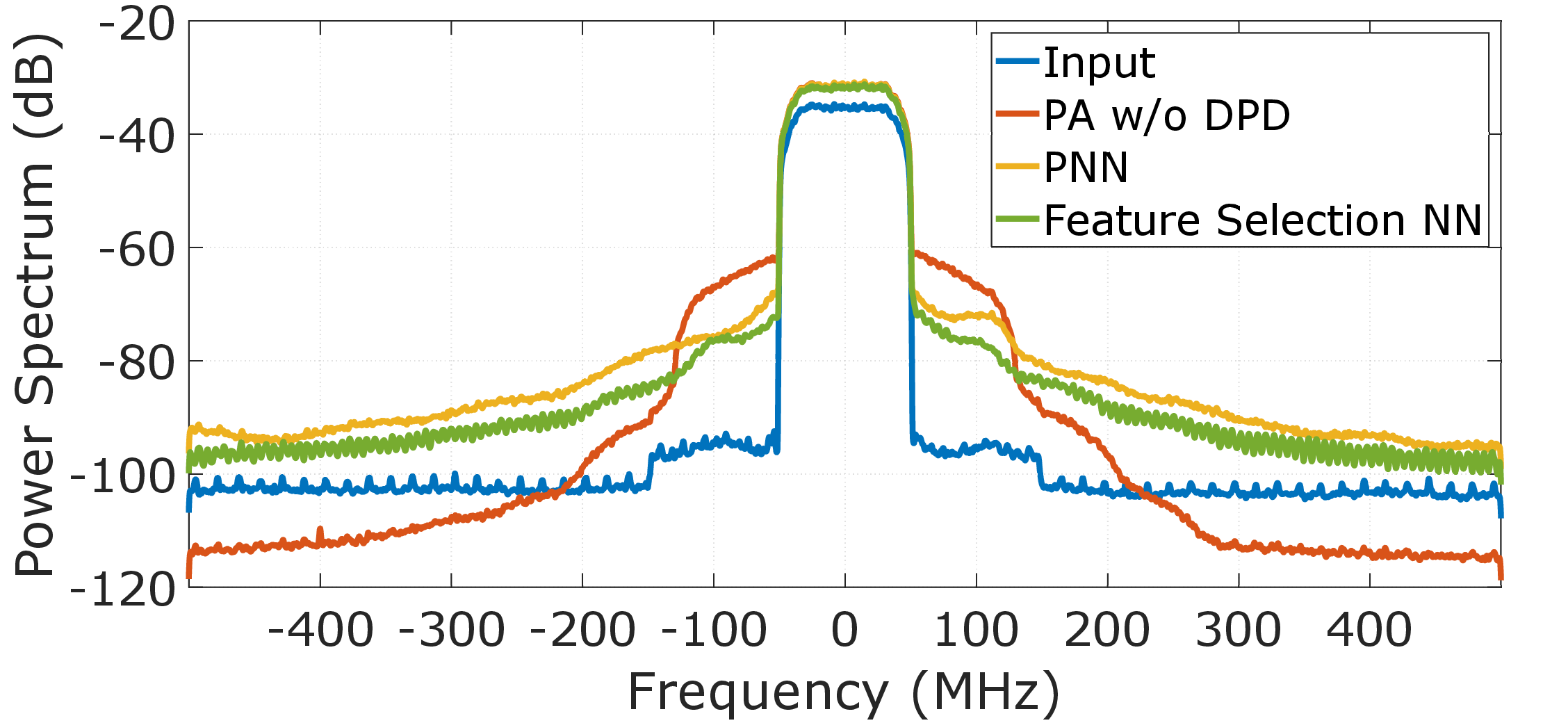}
\caption{Measured output spectrum after applying NN DPD algorithms with approximately 400 FLOPs.}
\vspace{-2mm}
\label{fig:measured_spectrum}
\end{figure}

\begin{figure}[!htbp]
        \centering
        \includegraphics[width=\linewidth, keepaspectratio]{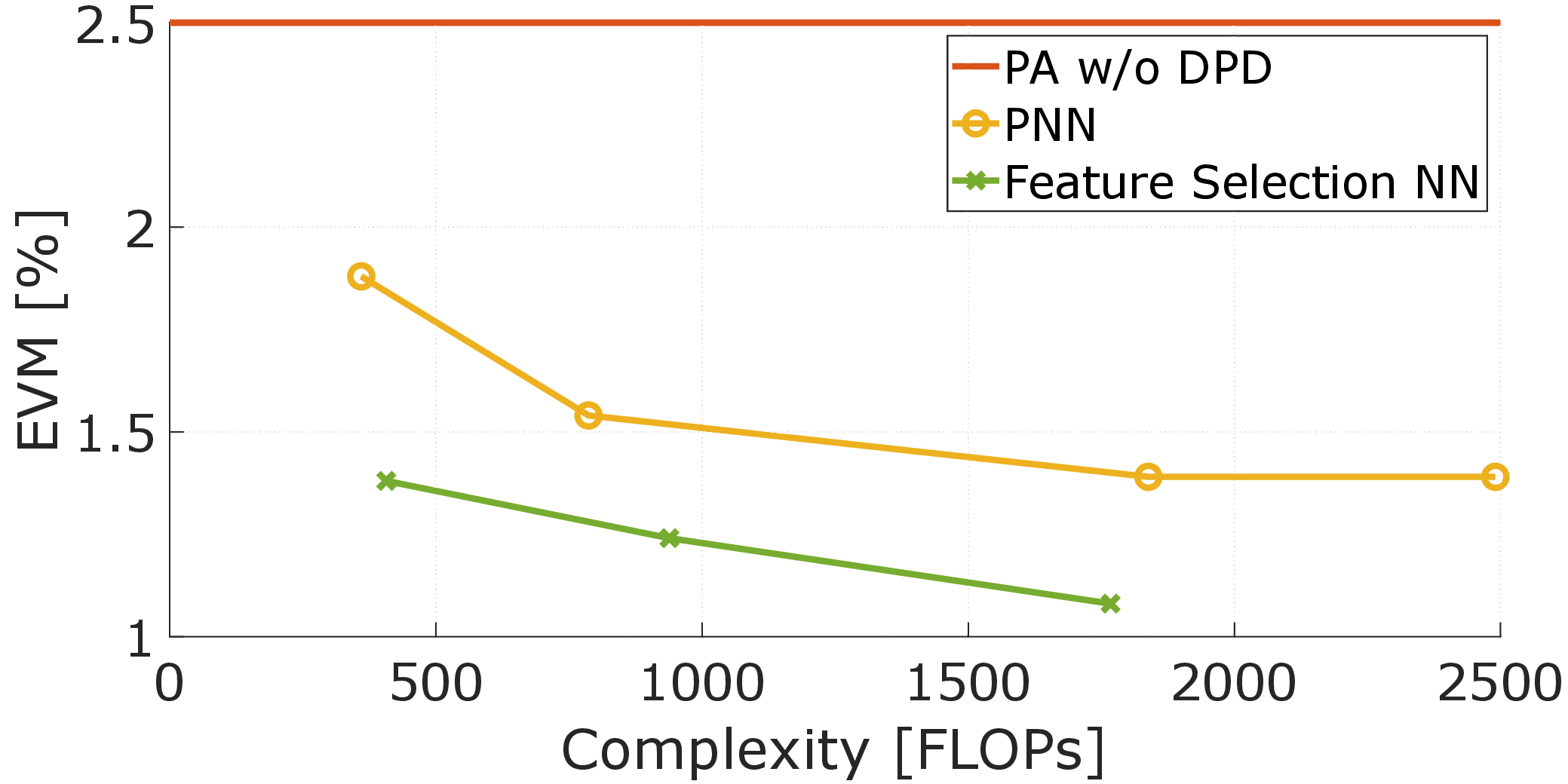}
    \caption{Measured EVM of different DPD algorithms.}
    \vspace{-3mm}
    \label{fig:measured_evm}
\end{figure}

Fig.~\ref{fig:measured_spectrum} shows the measured spectrum after DPD linearization using the PNN and proposed models with approximately 400 \ac{flops} (the corresponding FLOP numbers are highlighted in Table~\ref{table:measured_performance}). The proposed feature selection NN approach achieves the lowest ACLR in this comparison. Finally, Fig.~\ref{fig:measured_evm} shows the measured EVM as a function of model complexity, highlighting the excellent linearization capability of the proposed approach.
 
\section{Conclusion}  \label{sec:conclusion}
We proposed a Feature Selection Neural Network for low-complexity DPD operating in the FR3 frequency band, achieving an improved performance-complexity tradeoff compared to Volterra and Phase-normalized Neural Network baselines. Compared to the PNN baseline, the proposed approach reduces the computational cost by up to 30\%. The method is validated by measured EVM and ACLR performance results. An interesting topic for further work is the inspection of the selected features, possibly leading to more interpretable neural network DPD algorithms. The measured FR3 PA response datasets used to support this work are openly available at~\cite{thys_fr3_2026}.

\ifaccepted
\section*{Acknowledgment}
This work was supported by the 6G-MIRAI project, which has received funding from the Smart Networks and Services Joint Undertaking (SNS JU) under the European Union's Horizon Europe research and innovation program under Grant Agreement No 101192369. Views and opinions expressed are, however, those of the author(s) only, and they do not necessarily reflect those of the European Union or the SNS JU (granting authority). Neither the European Union nor the granting authority can be held responsible for them. The work of Rodney Martinez Alonso is supported by the Research Foundation–Flanders (FWO) under Grant 1211926N. The resources and services used in this work were provided by the VSC (Flemish Supercomputer Center), funded by the Research Foundation - Flanders (FWO) and the Flemish Government.
\fi

\bibliography{references}
\bibliographystyle{IEEEtran}

\end{document}